\documentclass[11pt]{article}
\usepackage[abs]{overpic}
\usepackage{amsmath,amssymb,amsfonts}
\usepackage{graphicx,color}
\usepackage{setspace} 
\usepackage{hhline}

\makeatletter

\@addtoreset{equation}{section}
\makeatother

\newcommand{\be}{\begin{equation}}

\newcommand{\ee}{\end{equation}}
\newcommand{\bea}{\begin{eqnarray}}
\newcommand{\eea}{\end{eqnarray}}
\newcommand{\beann}{\begin{eqnarray*}}
\newcommand{\eeann}{\end{eqnarray*}}

\newcommand{\ba}{\begin{array}}
\newcommand{\ea}{\end{array}}

\newcommand{\Tr}{\mathop{\rm Tr}}

\newcommand{\p}{{\partial}}
\newcommand{\N}{{\cal N}}
 
\def\XXint#1#2#3{{\setbox0=\hbox{$#1{#2#3}{\int}$} 
\vcenter{\hbox{$#2#3$}}\kern-.5\wd0}} 
 
\renewcommand{\thefootnote}{\fnsymbol{footnote}}

\begin{document}

\setlength{\oddsidemargin}{0cm}
\setlength{\baselineskip}{7mm}

\vfil\eject
\setcounter{footnote}{0}
\begin{flushright}
\normalsize
RIKEN-TH-190\\
\end{flushright}
    \begin{Large}  
       \begin{center}
         { \bf Uniformization, Calogero-Moser/Heun duality\\ and \\
Sutherland/bubbling pants
}
       \end{center}
        \end{Large}
        
\begin{center}
Ta-Sheng Tai\footnote[1]{
        e-mail address: tasheng@riken.jp
    }

\end{center}
\renewcommand{\thefootnote}{\arabic{footnote}}
\begin{small}
\begin{center}
{\it Theoretical Physics Laboratory, RIKEN,
                    Wako, Saitama 351-0198, JAPAN}
\end{center}
 \end{small}
\begin{abstract}
\noindent
{\normalsize 
Inspired by the work of Alday, Gaiotto and Tachikawa (AGT), 
we saw the revival of 
Poincar{\'{e}}'s uniformization problem 
and Fuchsian equations obtained thereof. 

Three distinguished 
aspects are possessed by Fuchsian equations. First, they are available via imposing a 
classical Liouville limit on level-two 
null-vector conditions. Second, they fall into 
some $A_1$-type integrable systems. Third, 
the stress-tensor present there (in terms of the $Q$-form) 
manifests itself as a kind of one-dimensional ``curve".

Thereby, a contact with the recently proposed Nekrasov-Shatashvili limit was soon made on the one hand, 
whilst the seemingly mysterious derivation of Seiberg-Witten prepotentials from integrable 
models become resolved on the other hand. 
Moreover, AGT conjecture can just be regarded as a quantum 
version of the previous Poincar{\'{e}}'s approach.

Equipped with these observations, we examined 
relations between spheric and 
toric (classical) conformal blocks via Calogero-Moser/Heun duality. 
Besides, as Sutherland model is also obtainable from 
Calogero-Moser by pinching tori at one point, we 
tried to understand its eigenstates 
from the viewpoint of toric diagrams with 
possibly many surface operators (toric branes) inserted. 
A picture called ``bubbling pants" then emerged and 
reproduced well-known results of the 
non-critical self-dual $c=1$ string theory 
under a ``blown-down" limit. 
}
\end{abstract}
\setstretch{1.1}

\section{Introduction }

Recently, AGT conjecture \cite{Alday:2009aq} 
has launched extremely active 
investigations towards both 2D Liouville CFT and 4D $\N=2$ 
$SU(2)$ SCFT or even generalization, say, 
higher rank of gauge groups \cite{Wyllard:2009hg}, 
non-conformal (asymptotically-free) limit 
\cite{Gaiotto:2009ma,Marshakov:2009gn} and 
$q$-deformation \cite{Awata:2009ur,Awata:2010yy} to name a few. 
Now, it seems rather appropriate to consider not only 
proving (or checking) this conjecture but also why 
these two apparently 
irrelevant arenas should coincide with each other.

In this short article, we observe that the 
``uniformization problem" pioneered by 
Klein, Koebe and Poincar{\'{e}} more than a century ago 
may shed illuminating 
light on answering 
this question%
\footnote{See 
\cite{Hadasz:2005gk, Hadasz:2006rb} 
for issues about 
uniformizing four-punctured Riemann spheres}. For instance, 
the stress-tensor $T(z)$ showing 
up there turns out to provide the $SU(2)$ Gaiotto (rewritten Seiberg-Witten) 
curve \cite{Gaiotto:2009we} under certain proper limit, 
say, large intermediate $s$-channel momentum. Consequently, 
AGT conjecture strongly manifests itself 
as a full 
quantum uplift of the uniformization problem w.r.t. two 
equivariant parameters $(\epsilon_{1},\epsilon_{2})$. 
In addition, Fuchsian type 
differential equations%
\footnote{Second-order differential equations are 
of Fuchsian type if they are fixed by certain Riemann scheme 
and meanwhile all their 
singularities are regular.} 
\begin{eqnarray}
\Big(\p_z^2 + T(z)\Big) \Psi=0 
\label{f1}
\end{eqnarray}
which are present during 
uniformizing punctured Riemann surfaces 
are as well available through the (semi)classical Liouville limit $b\to 0$ (or infinitely large central charge 
$c\to \infty$) 
imposed onto the constraint for a null vector at the second level 
in Verma module, i.e.  
\begin{eqnarray}
\label{vv}
\Big( L_{-2} -\frac{3}{2\big( 
2\Delta(h_{2,1}) +1 \big)}L^2_{-1} \Big) 
\Phi_{2,1}=0 , 
~~~~~~  h_{r,s}=\frac{1-r}{2}b + 
\frac{1-s}{2b},~~~~~~ h_{2,1}=-\dfrac{b}{2}.
\end{eqnarray}
As a matter of fact, this classical limit can get 
readily identified with 
the so-called 
Nekrasov-Shatashvili (NS) limit \cite{Nekrasov:2009rc} 
($g_s$: string coupling) 
\begin{eqnarray}
\label{NS}
\epsilon_1=b g_s\to0,
~~~~~~\epsilon_2=\frac{g_s}{b}=\text{fixed},
~~~~~~g_s\to0, ~~~~~~b\to 0
\end{eqnarray} 
which ultimately has something to do with many 
well-known integrable 
systems. Performing the WKB method onto the equation 
\eqref{f1} 
satisfied by $\N=2$ instanton 
partition functions with one surface operator inserted under NS limit, 
one is able to obtain either $classical$ 
conformal blocks \cite{Zamolodchikov:1995aa}%
\footnote{Their quantum versions are called 
Belavin-Polyakov-Zamolodchikov (BPZ) conformal blocks 
${\cal B}$ \cite{BPZ}. See the definition around \eqref{ccb}. } 
or 
$\log Z_{inst}(\epsilon_1, \epsilon_2=0)$ 
following AGT who claimed a much more general relationship \begin{eqnarray*}
Z_{inst}(\epsilon_1, \epsilon_2)={\cal B}(b, \frac{1}{b}).
\end{eqnarray*}
Many publications \cite{Alday:2009fs, 
Mironov:2009uv, Kozcaz:2010af,Alday:2010vg, Dimofte:2010tz,Maruyoshi:2010iu, Taki:2010bj, 
Awata:2010bz,Kozcaz:2010yp} have been devoted into this direction.

In summary, the philosophy we are after is as follows. 
Because every 
$\N=2$ $A_1$-type SCFT is associated with some genus-$g$ 
$n$-punctured Riemann surface $C_{g,n}$ \cite{Gaiotto:2009we}, 
through 
uniformizing $C_{g,n}$ both a classical piece of 
Liouville theory and an integrable (Fuchsian) system 
will be further inferred. This viewpoint therefore renders us the 
clue for understanding why these 
two seemingly different arenas ultimately 
meet.

Based on these, we start off to consider a four-punctured Riemann sphere ${\mathbb P}^1 \backslash \{z_1,\cdots,z_4 \}$ over which 
a second-order 
Heun equation is defined. 
Obviously, it can be thought of as an extension of usual hypergeometric differential equations living on 
${\mathbb C} \backslash \{0,1 \}$ fixed by certain 
Riemann scheme%
\footnote{See \cite{Tai:2010im} for an 
interpretation of triality 
in the $SU(2)$ Seiberg-Witten theory in terms of Gauss hypergeometric functions.}. 
\begin{table}[h]
\caption{Riemann scheme in the presence of five regular singular points on ${\mathbb P}^1$. $\chi$'s are called characteristic exponents. }
  \begin{center}
    \begin{tabular}{|c|c|c|c|c|} \hline
$z=0$& $z=1$& $z=t$&$z=\lambda$ &$z=\infty$ \\ \hline
0&0&0&0&$\chi_1$\\ \hline
$\chi_2$&$\chi_3$&$\chi_4$&$\chi_5$&$\chi_6$\\ \hline
    \end{tabular}
  \end{center}
\end{table}
It then seems 
plausible that classical spheric 
five-point conformal blocks having 
one null vector $\Phi_{2,1}$ insertion at the second 
level satisfy 
Heun equations. We clarify this observation via a 
celebrated mathematical duality: Heun can be 
suitably transformed into a two-body ($A_1$-type) elliptic Calogero-Moser (eCM) model. 
Finally, in view of a marvelous 
limit bringing eCM to Sutherland 
(trigonometric Calogero-Sutherland) model, 
we develop some correspondence between its excited 
wave functions and the toric diagram geometrically engineering 
${\cal T}_{0,3}(A_1)$ theory.

We organize this article as follows. 
In section 2, we explicitly re-derive the 
$2N_c=N_f=4$ SW prepotential 
upon 
taking into account two approaches: 
hermitian matrix model 
and classical Liouville theory. 
One may think of this presentation as another 
derivation other than 
the original AGT proposal. 
This is because we have just relied on 
the Ward identity of $T(z)$ 
in the uniformization problem 
and arguments intrinsic to hermitian matrix models. 
In section 3, we proceed to review Heun/eCM duality and 
study the relationship between toric and spheric conformal blocks 
(elliptic/non-elliptic $\N = 2$ 
$SU(2)\times SU(2)$ SCFTs). Also, we find new insights 
into eigenstates of 
Sutherland model from the toric diagram perspective. 
Finally, a chart summarizing the main idea is pasted in 
section 4.

\section{Uniformization problem}

Owing to Klein, Koebe and Poincar{\'{e}}, one is 
capable of uniformizing a punctured 
Riemann surface by means of 
hyperbolic geometry. For instance, 
there exists uniquely 
a hyperbolic 
metric 
\begin{eqnarray}
\label{1}
ds^2=e^{\varphi(z, \bar{z})} dz d\bar{z}
\end{eqnarray}
on an $n$-punctured 
Riemann sphere $X={\mathbb P}^1 \backslash
\{ z_1, \dots, z_n\}$ which 
has the constant negative curvature $R=-8\pi \mu b^2$ and is obviously 
in the same 
conformal class as the flat space $ds^2=dzd\bar{z}$. 
One is as well able to think of \eqref{1} as the pull-back from 
Poincar{\'{e}}'s hyperbolic metric of a unit disc ${\mathbf D}$ 
(or upper half-plane ${\mathbb H}$)%
\footnote{While what is isomorphic to $X$ is often 
${\bf D}/\Gamma$ with $\Gamma\subset PSU(1,1)$ being the 
isometry group of ${\bf D}$, it is harmless to directly 
make use of the metric of ${\bf D}$ in this manner.}
, i.e. 
\begin{eqnarray*}
\label{2}
ds^2=e^{\varphi(z, \bar{z})} dz d\bar{z}=
\frac{4| \eta' |^2}{\big( 1-|\eta|^2 \big)^2}
 dz d\bar{z}, 
 ~ ~ ~ ~ ~ ~ ~ ~ ~ ~
\eta(z):~\text{uniformization~map}. 
\end{eqnarray*}
Now, the requirement $R=-8\pi \mu b^2$ leads to 
the familiar Liouville equation 
\begin{eqnarray}
\label{3}
\partial_z \partial_{\bar{z}} \varphi =
2\pi \mu b^2 e^\varphi 
\end{eqnarray}
which is solved by $\varphi_{cl}$, the stationary point of some 
Liouville 
functional ${\cal S}[\varphi]$ being clarified later. 
Note that our conventions are listed below
\begin{eqnarray*}
Q=b+{1}/{b}, ~~~~
b:~\text{dimensionless~Liouville~coupling~constant}, 
 ~~~~\mu:~\text{cosmological~ constant}.
\end{eqnarray*}
Of course, the asymptotic behavior of $\varphi_{cl}$ 
around $z_i$ $(i=1,\dots,n)$ 
is readily read off according to 
the weight assigned to them. 
A guideline laid down by Polyakov to which 
we will adhere is that a path integral 
over $\phi$, say, 
\begin{eqnarray*}
\int {\cal D}\phi 
\prod_{i=1}^n V_i (z_i)
\exp({\cal S}[\phi])
\end{eqnarray*}
with 
$V_i=\exp(2\alpha_i \phi)$ 
whose conformal dimension is $\Delta_i=\alpha_i (Q-\alpha_i )$, should be 
equivalent to 
$\int {\cal D}\phi \exp({\cal S}[\phi])$ 
where instead $\phi$ is endowed 
with singularities at $z_i$'s and vertex operators become absent thereof. 
Set 
$\varphi={2b}\phi$ and 
$S[\phi]=\dfrac{1}{b^2}{\cal S}[\varphi]$ 
in order to facilitate a 
classical analysis through $b\to0$. Here, 
$S$ ($\phi$) stands for the $bare$ 
Liouville functional (field).

Next, 
construct from $\varphi_{cl}$ a standard Liouville 
$(2,0)$ stress-tensor $T_L (z)=
 {Q}\partial^2_z \phi
-(\partial_{{z}} \phi)^2$. 
According to 
Poincar{\'{e}} one has the following 
expansion (when all punctures 
are $parabolic$):  
\begin{eqnarray}
\label{6}
T(z)\equiv 
\frac{1}{2}\partial^2_z \varphi_{cl}
-\frac{1}{4}(\partial_{{z}} \varphi_{cl})^2
=\sum_{i=1}^{n-1} \frac{1}{4(z-z_i)^2} + \frac{c_i}{(z-z_i)}
\end{eqnarray}
where $accessory$ parameters $c_i$'s satisfy three linear constraints yielded 
by imposing 
\begin{eqnarray*}
\label{}
T(z)\to \frac{1}{4 z^2} + \frac{c_n}{z^3}+{\cal O}\Big(
\frac{1}{z^4}\Big),  ~~~~~~~~~~z\to \infty.
\end{eqnarray*}
We will 
always assume $(z_{1}, 
z_{n\text{-}1}, z_n)=(0,1,\infty)$ under 
a suitable M{\"o}bius transformation. 
Note that accessory parameters can get 
either determined by 
monodromies of \eqref{f1} on ${\mathbb P}^1$ or the so-called 
Polyakov conjecture upon exploiting the 
Ward identity w.r.t. $T_L (z)$: 
\begin{eqnarray}
c_i=-\frac{\partial {\cal S}[\varphi_{cl}(\delta_i,z_i)]}{\partial z_i}, 
 ~~~~~~~~~i\ne(0,1,\infty).
 \label{po}
\end{eqnarray}
In contrast to $\Delta_i$, here 
$\delta_i=\xi_i (1-\xi_i)
=\frac{1}{4}(1-\mu_i^2)$  
denotes the $classical$ conformal weight of inserted vertex 
operators. Besides, each 
zero (non-zero) $\mu_i$ corresponds to a parabolic (elliptic) puncture. Remarkably, 
\eqref{po} has been proved rigorously by Zograf and Takhtajan \cite{ZT} 
when all punctures are parabolic ones.

\subsection{Fuchsian differential equation}
Recall that the $Q$-form of 
Fuchsian equations looks like 
\begin{eqnarray}
\partial_z^2 y + \dfrac{1}{2}\{ \rho ,z \}y=0, ~~~~~~
T(z)= \dfrac{1}{2} \{ \rho ,z \}:~ \text {Schwarzian derivative of}~ \rho, ~~~~~~\rho=\dfrac{y_\vartheta}{y_\varsigma}.
\label{ft}
\end{eqnarray}
\eqref{ft} is fixed by both 
the number of regular singularities placed 
on ${\mathbb C}$ and a prescribed Riemann scheme. 
Here, the 
multivaluedness of $\rho$ 
is accounted for due to its form being 
a ratio of two independent solutions to \eqref{ft}. 
Also, the pair $(y_\vartheta, y_\varsigma)$ may get normalized to have a unit Wronskian: 
\begin{eqnarray*}
(\partial_z y_\vartheta \cdot y_\varsigma - y_\vartheta \cdot 
\partial_z y_\varsigma)=1
\end{eqnarray*} 
which helps fix the conjugation acquired 
from $SL(2,{\mathbb C})$. 
In view of \eqref{ft}, a type of 
Riemann-Hilbert (RH) problem can be raised. Namely, 
once monodromies of $\rho$ in $SU(1,1)$ are found 
explicitly, all $c_i$'s get known 
subsequently. This sounds like 
the usual RH problem only when the latter statement 
is 
replaced by ``there exists certain algebraic curve whose 
period integrals reproduce $(y_\vartheta, y_\varsigma)$".

The stress-tensor $T(z)$ thus obtained by computing 
Schwarzian is a meromorphic function on 
${\mathbb P}^1$: 
\begin{eqnarray}
\label{me}
T(z)=\sum_{i=1}^{n-1} \frac{\delta_i}{(z-z_i)^2} + \frac{c_i}{(z-z_i)}. 
\end{eqnarray}
As before, its asymptotic behavior at infinity is 
supposed to be regular such that 
\begin{eqnarray*}
T(z)\to \frac{\delta_n}{ z^2} + \frac{c_n}{z^3}+{\cal O}\Big(
\frac{1}{z^4}\Big)
\end{eqnarray*}
is able to impose three linear equations on $c_i$'s: 
\begin{eqnarray*}
\sum_{i=1}^{n-1} c_i=0, ~~~~~~~~
\sum_{i=1}^{n-1}(\delta_i + c_i z_i)=\delta_n, ~~~~~~~~
\sum_{i=1}^{n-1}(2\delta_i z_i + c_i z_i^2)=c_n.
\end{eqnarray*}
As a remark, according to 
\begin{eqnarray*}
\label{emt}
T(z)=\frac{1}{2}\{ \rho(z), z \}=
\frac{2 \partial^3 \rho \cdot \partial \rho 
- 3(\partial^2 \rho)^2}
{4(\partial \rho)^2},~~~~~~~~\partial \equiv \partial_z,
\end{eqnarray*}
an equivalent description of this expression goes back precisely to 
\eqref{2} if one equates $\eta$ with $\rho$.

Let us illustrate more concretely what was announced 
around 
\eqref{vv}. That is, define 
\begin{equation*}
\label{chi}
\left\langle\Phi_{2,1}(z) \right\rangle 
\equiv \Psi(z)=
\Big\langle
V_4(\infty)V_3(1)
\Phi_{2,1}(z) V_2(x)V_1(0) \Big\rangle
\end{equation*}
at the level of BPZ conformal blocks. 
\eqref{vv} lays down ($\Delta=h_{2,1}$)
\begin{eqnarray*}
\label{Fuchs1}
&&\left[
\frac{\partial^2}{\partial z^2}
+\gamma\left(\frac{1}{z} - \frac{1}{1-z}\right)\frac{\partial}{\partial z}-\gamma
\Big(\frac{\Delta_1}{z^2} + \frac{\Delta_2}{(z-x)^2}
+  \frac{\Delta_3}{(1-z)^2} +
\frac{\Lambda_{}}{z(1-z)}
+ \frac{x(1-x)}{z(z-x)(1-z)}
\frac{\partial}{\partial x}\Big)
\right]\\
&&\times
\left\langle\Phi_{2,1}(z) \right\rangle =0
\end{eqnarray*}
for the null-vector where 
\begin{eqnarray*}
\Lambda  =
\Delta_1 + \Delta_2 + \Delta_3 + \Delta - \Delta_4, 
~~~~~~~~~~~
\gamma= \frac23(2\Delta + 1).
\end{eqnarray*}
By $b\to0$, 
the above equation becomes Fuchsian: 
\begin{equation*}
\Big(\partial^2_z +T(z) \Big) \Psi^{cl}(z) = 0
\end{equation*}
where 
\begin{eqnarray*}
T(z)=\frac{\delta_1}{z^2} + \frac{\delta_2}{(z-x)^2}
+  \frac{\delta_3}{(1-z)^2}
+ \frac{\delta_1+\delta_2 + \delta_3 - \delta_4}{z(1-z)}
+ \frac{x(1-x){c}(x)}{z(z-x)(1-z)}.
\end{eqnarray*}
$T(z)$ here 
does not look like \eqref{me} 
because 
$(0,1,\infty)$ have been chosen in order to eliminate constraints due to projective invariance. Note that the 
accessory parameter $c(x)$ will be given in \eqref{pppp} later on.

\subsection{Classical Liouville theory}
From now on, we mainly focus on ${\mathbb P}^1 
\backslash \{ 0,x,1,\infty \}$ such that 
$z_2=x$ stands for the cross-ratio. 
While the classical limit $b\to 0$ is taken, 
an $n$-point function ${\cal G}(\alpha_1, \cdots, 
\alpha_n)$ computed w.r.t. $S[\phi]$ gets approximated by   
\begin{eqnarray*}
{\cal G}(\alpha_1, \cdots, 
\alpha_n)\sim \exp\Big(-\dfrac{1}{b^2}{\cal S}[\varphi_{cl}]\Big)
=\exp\Big(-\dfrac{1}{b^2}{\cal S}_{cl}[\xi_1, \cdots, 
\xi_n; z_1, \cdots, z_n]\Big) 
\end{eqnarray*}
where $\varphi_{cl}$ again satisfies \eqref{3} as well as 
all required asymptotics with $\alpha_i=\xi_i/b$ being very heavy. 
Moreover, as $Q^2 \to {1}/{b^2}$ the quantum BPZ conformal block becomes 
\begin{eqnarray}
{\cal F}_{ \Delta}\left[
\begin{array}{cc}
\Delta_3&\Delta_2  \\
\Delta_4&\Delta_1\\
\end{array}
\right](x) \sim 
\exp\Big(Q^2 
f_{\delta}\left[
\begin{array}{cc}
\delta_3&\delta_2  \\
\delta_4&\delta_1\\
\end{array}
\right](x)
\Big), ~~~~~~~~\delta=\frac{1}{4}+p^2
\label{ccb}
\end{eqnarray}
where $f$ represents 
the classical conformal block \cite{Zamolodchikov:1995aa}. 
Also, since $Q^2\to \infty$ the intermediate momentum $p$ is 
forced to be $p_s$, $s$-channel 
saddle point momentum, which solves 
($\delta_s \equiv \frac{1}{4}+p_s^2$)
\begin{eqnarray*}
\frac{\partial {\cal S}[\xi_1,\cdots,\xi_4; \delta; x]}{\partial p}\Big|_{p=p_s} =0 , ~~~~~~~~~
{\cal S}[\xi_1,\cdots,\xi_4; \delta_s; x]=
{\cal S}_{cl}[\xi_1,\cdots,\xi_4;  x].
\end{eqnarray*}
Needless to say, ${\cal F}_{\Delta}(\Delta_i,x )$ 
plays a central rule in the full Liouville 
four-point function%
\footnote{We have omitted the factor 
$\prod_{i<j}|x_{ij}|^{2\gamma_{ij}}$ with 
$\gamma_{12}=\gamma_{13}=0, 
\gamma_{14}=-2\Delta_{1}, 
\gamma_{24}=\Delta_{1}+\Delta_{3}-\Delta_{2}-\Delta_{4}, 
\gamma_{34}= \Delta_{1}+\Delta_{2}-\Delta_{3}-\Delta_{4}~\text{and}~  
\gamma_{23}=\Delta_{4}-\Delta_{1}-\Delta_{2}-\Delta_{3}$ 
inside ${\cal G}$.}
: 
\begin{eqnarray}
{\cal G}(\alpha_1, \cdots, 
\alpha_4; x, \bar{x})=\frac{1}{2}\int^{\infty}_{-\infty}dP~
C(\alpha_1,\alpha_2,\frac{Q}{2}+iP)\cdot
C(\alpha_3,\alpha_4,\frac{Q}{2}-iP)\cdot
\big|{\cal F}_{\Delta}(\Delta_i;x )\big|^2 
\label{full}
\end{eqnarray}
where $C$ is recognized as the structure constant and 
$P=p/b$ ($P^2=\Delta-\frac{1}{4}Q^2$).

What comes as a classical counterpart of \eqref{full} is 
\begin{eqnarray}
{\cal S}_{cl}[\xi_1,\cdots,\xi_4;  x]
={\cal S}^{(3)}(\delta_4, \delta_3 ,\delta_s)
+{\cal S}^{(3)}(\delta_s, \delta_2 ,\delta_1)
-f_{\delta_s,\delta_i}(x) -{f}_{\delta_s, \delta_i}(\bar{x}).
\label{cfull}
\end{eqnarray}
What was referred to as Polyakov conjecture now states that 
\begin{eqnarray}
\label{pppp}
c_2(x)=
-\frac{\partial {\cal S}_{cl}[
\xi_i;x ]}{\partial x}=
\Big( \frac{\partial }{\partial x} 
f_{\delta}\left[
\begin{array}{cc}
\delta_3&\delta_2  \\
\delta_4&\delta_1\\
\end{array}
\right](x)\Big)_{p=p_s(x)} 
\end{eqnarray}
where the second equality is derived by taking into account \eqref{cfull}. 
Notice that $\ell(\gamma_{12})= 4\pi p_s$ where 
$\gamma_{12}\equiv\gamma(x)$ 
represents the geodesic length 
of a (hyperbolic) four-punctured Riemann 
sphere and encircles two punctures $(z_1,z_2)\equiv(0,x)$ 
from other ones.

\subsection{Hermitian matrix model}

Because the relationship amongst \eqref{pppp} is 
entirely chiral and stems from 
the familiar Ward identity about $T_L (z)$ inserted at the level 
of conformal blocks, it is tempting to compare \eqref{pppp} 
with what takes place in a hermitian matrix model 
possessing an usual Vandermonde%
\footnote{This section is fully inspired by (3.41) of 
Eguchi and Maruyoshi \cite{Eguchi:2010rf}. }. 
That is, by means of Kodaira-Spencer field 
(chiral free boson) $\phi_{KS}$, 
the expectation value of the matrix model 
stress-tensor $T_M (z)$ given by 
\begin{eqnarray*}
&&y^2=\langle T_M(z) \rangle\to
{\cal W}' (z)^2  +f(z),~~~~~~~~~~~~\text{when}~\hbar=1/N \to 0,\nonumber\\
&&T_M (z)=\sum_{n} L_{n} z^{-n-2} =(\partial \phi_{KS})^2,\nonumber\\
&&Z=\frac{1}{\text{vol}U(N)}\int_{N\times N} dM \exp \Big( 
\frac{1}{\hbar} \Tr{\cal{W}} (M)\Big)
=\exp \Big(\sum_{g\ge 0} \hbar^{2g-2} {\cal{F}}_g \Big),\nonumber\\
&&\langle \p \phi_{KS}(z) 
\rangle={\cal W}'(z) + 2\hbar\Tr\log\big(z-M \big)
=
\Big( t_0 z^{-1}+\sum_{n>0} n t_n z^{n-1} 
+2\hbar^2 \sum_{n\ge 0} z^{-n-1}\frac{\p}{\p t_n} \Big)Z
\end{eqnarray*}
leads to the large-$N$ spectral curve. 
Certainly, 
its Ward identity bears a strong resemblance to 
\eqref{pppp} but now $f$ has to be replaced by 
$\hbar^{-2}{\cal F}_0$ (with possibly $\phi_{KS} \to \hbar^{-1}\phi_{KS}$ as $\hbar\to 0$). 
$L_n$'s are Virasoro generators realized 
by $\phi_{KS}$ possessing the central charge $c=1$.

We naturally anticipate that if the spectral curve 
$y^2=\lim_{\hbar\to 0}\langle T_M(z)\rangle$ 
is recognized as Gaiotto curve (rewritten Seiberg-Witten 
curve), ${\cal F}_0$ of $Z$ gets equivalent to 
the low-energy $\N=2$ SW 
prepotential. The concrete form of $Z$ has already been 
proposed by Dijkgraaf and Vafa \cite{Dijkgraaf:2009pc} 
last year (see \cite{Itoyama:2009sc, Eguchi:2009gf, Fujita:2009gf} for further refinements). 
Below, we will see that 
$f_{\delta_s, \delta_i}(x)$ of \eqref{ccb} 
turns out to give us 
the desired infra-red prepotential ${\cal F}^{SW}$.

\subsection{From 
null-state condition to Fuchsian (Schr\"{o}dinger-like) equation}

Another distinguished aspect we 
want to review involves 
a degenerate field $V_{-\frac{b}{2}}$ entering 
the standard Liouville theory. 
Due to the 
null-vector decoupling equation at 
the second level $L^2_{-1}+b^2 L_{-2}=0$, 
conformal blocks involving $V_{-\frac{b}{2}}$ 
obey 
\begin{eqnarray}
b^{-2}\frac{\p^2}{\p z^2} + 
\sum_{i=1}^{n-1} \Big(\frac{\Delta_i}{(z-z_i)^2} + 
\frac{1}{(z-z_i)}\frac{\p}{\p z_i}\Big) \big\langle 
V_{-\frac{b}{2}}(z)\prod_{i=1}^{n-1}V_i(z_i)\big\rangle=0
\label{ddddd}
\end{eqnarray}
Taking the (semi)classical limit $b\to 0$ in \eqref{ddddd} 
ultimately recovers \eqref{ft} as done around Section 2.1. 
Of course, a direct analogy can soon be seen 
in hermitian matrix models upon using the $c=1$ 
$\phi_{KS}(z)$; namely, 
the vev of a determinant operator 
\begin{eqnarray}
\big\langle \det (z-M)\big\rangle
=\big\langle \exp(\frac{1}{\hbar}\phi_{KS})\big\rangle=
\exp\Big(
\frac{1}{\hbar}\int^z dy ~\langle \p \phi_{KS}(y)\rangle +{\cal O}(\hbar^0)\Big)
\label{siki}
\end{eqnarray}  
has been known to solve the Schr\"{o}dinger-like 
equation stemming from 
the genus-zero spectral curve and meanwhile serves as 
the orthogonal polynomial for the matrix model. For example, 
\eqref{siki} may stand for $H_N(z)$ 
(Hermite polynomial) when $Z$ is Gaussian.

\subsubsection{Reproducing Seiberg-Witten prepotential}
Let us proceed to work out some examples explicitly 
in which 
$2N_c=N_f=4$ Seiberg-Witten prepotentials are recovered 
upon employing known 
classical 4pt-spheric conformal blocks. 
Extracting SW prepotentials from classical conformal blocks 
may be viewed as another derivation 
more or less independent of the original AGT proposal. 
This is 
because we have just taken advantage of both the Ward identity 
of the stress-tensor intrinsic to the 
uniformization problem (Polyakov conjecture) and arguments 
familiar in large-$N$ hermitian matrix models.

Here, bare flavor masses and weights assigned to 
punctures are related by  
\begin{eqnarray*}
\begin{cases}
\xi_1=m_1 +m_2 + \frac{1}{2},~~~~~~~
\xi_2=-m_1+m_2 +\frac{1}{2},\\
\xi_3=m_3 +m_4 + \frac{1}{2},
~~~~~~~\xi_4=-m_3+m_4 +\frac{1}{2},
\end{cases}
\end{eqnarray*}
which descends directly from the so-called AGT dictionary 
\cite{Alday:2009aq}. Besides, 
in \cite{Teschner:2010je} Teschner pointed out that  
\eqref{ft} is referred to as the 
``Baxter equation" and applying 
to it the WKB approximation at the zeroth-order 
(much resembling \eqref{siki}) gives 
\begin{eqnarray}
a\equiv \oint dz ~\sqrt{T(z)}= 
\frac{\ell(\gamma_{12}) }{4\pi b} 
~~~~~~~~\text{or} ~~~~~~~~ p_s=\frac{a}{b}.
\label{swa}
\end{eqnarray}
Notice that $a$ expressed 
in terms of a period integral of 
$\sqrt{T(z)}$ coincides with the $SU(2)$ Coulomb phase parameter 
as claimed in \cite{Gaiotto:2009we, Alday:2009aq}. 
Equipped with these, we first examine the case of four massless flavors. 
 
\subsection*{(I) All $m_i=0$}
In view of \eqref{swa}, 
under the large-$p_s$ limit implemented by 
$b\to 0$, $f_{\delta_s, \delta_i}(x)$ $(\forall \delta_i=\frac{1}{4})$ one has 
\begin{eqnarray}
\label{mann}
&&f_{\delta_s, \delta_i =\frac{1}{4}} (x)\nonumber\\
&&=(p_s^2 -\frac{1}{4})\log x + (p_s^2 +\frac{1}{4})\frac{x}{2}
+ (\frac{13p_s^2}{16}+\frac{9}{32}+\frac{1}{256p_s^2 +256})\frac{x^2}{4}
+\cdots\nonumber\\
&&\to \frac{a^2}{b^2} (\log x + \frac{x}{2} +\frac{13 x^2}{64}
+\frac{23 x^3}{192} +\cdots )\equiv\frac{1}{b^2}
{\cal F}^{SW}_{inst}
\end{eqnarray} 
where in the second line we have borrowed 
(8.20) of Zamolodchikov and Zamolodchikov \cite{Zamolodchikov:1995aa}. 
Notice that the last line is precisely 
the desired SW prepotential (up to a perturbative piece 
$-\log 16$) 
via $b\equiv\hbar$ as well as $x\equiv \exp 
(2\pi i \tau_{UV})$. 

\subsection*{(II) $m_1=m_2=\frac{\xi}{4}$ and $m_3=m_4=0$}
Let us quote from \cite{Hadasz:2006rb} the following 
classical conformal block 
\begin{eqnarray}
\label{xi}
{f}_{\frac{1}{4} +p^2}\left[
\begin{array}{cc}
\frac{1}{4}&\frac{1-\xi^2}{4}  \\
\frac{1}{4}&\frac{1}{4}\\
\end{array}
\right](x) &&=
(p^2 -  \frac{1-\xi^2}{4})\log x +
(\frac{1-\xi^2}{8} + \frac{p^2}{2})x \nonumber\\
&&+\Big(\frac{9(1-\xi^2)}{128} + 
\frac{13p^2}{64} +   \frac{(1-\xi^2)^2}{1024(1+p^2)}
\Big)x^2 + {\cal O}(x^3).
\end{eqnarray}
Again, by $b\to0$ which leads to 
both large-$p$ and large-$\xi$ limits, we finally 
observe 
that \eqref{xi} does correctly reproduce the 
instanton part of SW prepotential for 
$\vec{m}=(m,m,0,0)$ 
\cite{Marshakov:2009kj}: 
\begin{eqnarray*}
{\cal F}_{inst}^{SW}=(a^2 -m^2)\log x+ (a^2 + m^2)\frac{x}{2}
+\Big( 13a^2 + 18 m^2 
+\frac{m^4}{a^2}+{\cal O}(a^{-4}) \Big)\frac{x^2}{64} +{\cal O}(x^{3})
\end{eqnarray*}
through carrying out $(a^2, m^2) \to(p^2, -\frac{\xi^2}{4})$.

\subsection*{(III) Arbitrary four flavor masses}
In order to deal with this case, we quote from 
\cite{Zamolodchikov:1995aa} the 
asymptotic expansion for generic $\delta_i$ and take 
both large $p$- and $\xi_i$-limit, i.e. 
\begin{eqnarray}
\label{em}
f_{\delta, \delta_i}(x)
&&=
(\delta-\delta_1-\delta_2)\log x + 
\frac{(\delta+\delta_1-\delta_2)
(\delta+\delta_3-\delta_4)}{2\delta}x + {\cal O}(x^2)\nonumber\\
&&\to (a^2 + 2m_1^2 + 2m_2^2)\log x + 
\frac{a^4 - 4a^2(m_1 m_2 + m_3 m_4)+16m_1 m_2 m_3 m_4}
{2a^2}x+\cdots
\end{eqnarray}
In the second line by $m_k \to \frac{i}{2}m_k$ one recovers 
the instanton part of $SU(2)$ (instead of $U(2)$) 
SW prepotential. For instance, it becomes 
\begin{eqnarray*}
(a^2 -m^2)\log x + 
\frac{(a^4 + 2a^2 m^2 + m^4)}{2a^2}x+\cdots
\end{eqnarray*}
for $m_1 =m_2 =m_3 =m_4=\frac{i}{2}m$ which agrees to 
(3.48) of \cite{Eguchi:2010rf} up to the $U(1)$ part 
$-2(m_1 +m_2)(m_3 +m_4)\log (1-x)$. 
Certainly, in order to get higher-order corrections 
one finally has to appeal 
to the BPZ representation of conformal blocks: 
\begin{eqnarray}
\label{zb}
&&{\cal F}_{\Delta}(\Delta_{1,2,3,4};x )
= x^{\frac{Q^2}{4}-\Delta_1-\Delta_2}
(1-x)^{\frac{Q^2}{4}-\Delta_1-\Delta_3}\nonumber\\
&&\times (16q)^{P^2}\big(\theta_3(q)\big)^{3{Q^2}-4(\Delta_1+
\Delta_2+\Delta_3+\Delta_4)} H_{\Delta}(\Delta_i ;q)
\end{eqnarray}
where $x$ and $q$ are related by the celebrated 
elliptic $lambda$-function:
\begin{eqnarray*}
x=\frac{\theta_2^4(q)}{\theta_3^4(q)}=16q-128q^2+704q^3-
3072q^4+11488q^5+\cdots
\end{eqnarray*}
with $q=\exp(i\pi \tau)$ and $\tau$ being called half-period. 
By first executing 
Zamolodchikov's recursive formula \cite{re} 
for $H=1+\sum_{n=1}^{\infty} H_{(n)} q^n$ and then 
taking 
$(\alpha_i, P)\gg Q$ inside \eqref{zb} 
(without the intermediate $b\to0$ 
procedure) one is able to obtain 
to any desired order the 
$2N_c=N_f=4$ SW prepotential.

\section{Application}
We have learned that dealing with 
uniformizing $C_{g,n}$ gets 
equivalent to 
studying Fuchsian 
equations \eqref{f1} 
whose solutions involve conformal blocks within which 
at least one $\Phi_{2,1}$ is inserted. 
A contact with well-known integrable 
systems (Heun, Sutherland, etc.) made by 
\eqref{f1} hopefully lays down opportunities for 
gaining insights into the corresponding $\N=2^\ast$ $SU(2)$ SCFT 
and ${\cal T}_{0,3}(A_1)$, a theory of four 
free hypermultiplets.

First of all, the solution to 
Heun equations must be 
interpreted as a spheric five-point  
conformal block ${\cal B}_5$%
\footnote{${\cal B}_n$ is an abbreviation for an 
$n$-point conformal block. } under $b\to 0$. Subsequently, by using 
Heun/eCM duality this ${\cal B}_5$ coincides with 
a toric ${\cal B}_2$ which as well as 
${\cal B}_5$ involves 
one $\Phi_{2,1}$ insertion. Thereafter, we consider 
a limit taking eCM to Sutherland model which 
corresponds to the operation of pinching a torus at one point. Additionally, its 
eigenstates are examined by making use of toric diagrams 
engineering responsible ${\cal T}_{0,3}(A_1)$ systems.

\subsection{Calogero-Moser/Heun duality}
We quickly review Calogero-Moser/Heun duality. 
Recall that Heun equation is of second-order: 
\begin{eqnarray}
\left(\left(\frac{d}{dz}\right)^2+\left(\frac{\gamma}{z}+\frac{\delta}{z-1}+\frac{\epsilon}{z-t}\right)\frac{d}{dz}
+\frac{\alpha\beta z -q}{z(z-1)(z-t)}\right) F(z)=0
\label{eq1}
\end{eqnarray}
with the constraint 
$\gamma+\delta+\epsilon=\alpha+\beta+1$.
This differential equation is of Fuchsian type, namely, 
all singularities $(0,1,t,\infty)$ have to be regular.

It should be emphasized that any second-order differential equation 
with four regular singularities on ${\mathbb P}^1$ falls into 
Heun upon suitable 
transformations. Due to 
\begin{eqnarray*}
{\wp}(x) ={\wp}(x|2\omega_1,2\omega_3)
=\frac{1}{x^2}+\sum_{(m,n)\in {\mathbb Z}\times{\mathbb Z} \backslash {(0,0)}}
\left(\frac{1}{(x-2 m \omega_1-2 n \omega_3)^2}
-\frac{1}{(2m\omega_1+2n\omega_3)^2}\right), 
\end{eqnarray*}
the fundamental period of Weierstrass function falls 
into $(2\omega_1,2\omega_3)$, i.e. 
$\wp(x+2\omega_1) =\wp(x+2\omega_3)=\wp(x)$. 
A torus ${\mathbb C}/(2\omega_1 {\mathbb Z}+2\omega_3 {\mathbb Z})$ thus comes from 
${\mathbb P}^1$ owning branching 
points $(\omega_0,\omega_1,\omega_2,\omega_3)$ with 
$\omega_0=0$ and $\omega_2=-\omega_1-\omega_3$.

Also, by setting 
\begin{eqnarray*}
e_i=\wp(\omega_i) , ~~~~t=\frac{e_3-e_1}{e_2-e_1}, ~~~~ z=\frac{\wp(x)-e_1}{e_2-e_1}, ~~~~ (i=1,2,3)
\label{eq5}
\end{eqnarray*}
four branching points are brought to 
$(0,1,t,\infty)$. 
Moreover, through $\eta(z) = z^{-\frac{l_0}{2}}(z-1)^{-\frac{l_1}{2}}(z-t)^{-\frac{l_2}{2}}$ and  $F(z)\eta(z)=f(x)$ 
Heun equation is transformed into
\begin{eqnarray}
\left( -\frac{d^2}{dx^2}+\sum_{i=0}^3 l_i(l_i+1)\wp (x+\omega_i)-E\right) f(x)=0 
\label{eq6}
\end{eqnarray}
where parameters are related by   
\begin{eqnarray}
&& l_0=\beta-\alpha -1/2, ~~~~l_1=-\gamma+1/2, ~~~~l_2=-\delta+1/2, ~~~~l_3=-\epsilon+1/2,\nonumber \\
&&E=(e_2-e_1)\Big(-4q+\big(-(\alpha-\beta)^2+2\gamma^2+6\gamma\epsilon+2\epsilon^2
-4\gamma-4\epsilon-\delta^2+2\delta+1\big)/3\nonumber\\
&& +\big(-(\alpha-\beta)^2+2\gamma^2+6\gamma\delta +2\delta^2-4\gamma-4\delta-\epsilon^2+2\epsilon+1\big)t/3\Big).
\label{eq7}
\end{eqnarray}
Note that 
transforming \eqref{eq1} into \eqref{eq6} is 
generally not unique. 
When $l_1=l_2=l_3=0$ $(\gamma=\delta=\epsilon=1/2)$, 
\eqref{eq6} is called 
Lam$\acute{\rm e}$ equation.

\subsection{Sphere versus torus}
Quite analogous to hypergeometric 
differential equations, Heun ones are designated 
to deal with 
four regular singularities placed on ${\mathbb P}^1$. Based on previous discussions, 
$\lim_{b\to 0}{\cal B}_5$ having one 
$\Phi_{2,1}$ insertion is naturally supposed to 
obey Heun equations. 
We will show that this is true upon using the result 
of Fateev, Litvinov, Neveu and Onofri \cite{Fateev:2009me} together with 
Heun/eCM duality.

In \cite{Fateev:2009me} ${\cal B}_5$ was written down as%
\footnote{For the sake of brevity, 
$V_i$ stands for $V_{\alpha_i}$.} 
\begin{eqnarray*}
&&{\cal B}_5\equiv \big\langle V_{-\frac{b}2}(z) V_1(0)V_2(1)V_3(\infty)V_4(x)
\big\rangle
=z^{\frac{1}{2b^2}}(z-1)^{\frac{1}{2b^2}}\\
&&\times \frac{\big(z(z-1)(z-x)\big)^{\frac{1}{4}}}{\big(x(x-1)\big)^{\frac{8\Delta_4 +1}{12}}}
\frac{\Theta_1(u)^{\frac{1}{b^2}}}
{\Theta'_1(0)^{\frac{1+b^2}{3b^2}}}
 \Phi(u|\tau)
\end{eqnarray*}
where
\begin{eqnarray}
\label{JJ}
&& \Big(-\partial_u^2+{V(u)} \Big)
\Phi(u|\tau)
=\frac{4i}{\pi b^2}\partial_\tau \Phi(u|\tau), ~~~~~~~~~
V(u)= \sum_{i=0}^3 s_i(s_i+1)\wp (u-\omega_i),\nonumber\\
&&u=\frac{\pi}{4K(x)}
\int^{(z-x)/(xz-x)}_{0} 
\frac{dt}{\sqrt{t(1-t)(1-xt)}}, ~~~~~~~~~
2\alpha_i=Q -b(s_i+\frac{1}2).
\end{eqnarray}
Notice that $\Theta_1(u)$ is Jacobi theta function whilst 
$K(x)$ denotes the complete elliptic integral of the 
first kind. By $b\to0$ one observes that 
\begin{eqnarray*}
&& \Big(-b^2\partial_u^2+{U(u)} \Big)
\Phi^{cl}(u|\tau)
=\frac{4i}{\pi}\partial_\tau \Phi^{cl}(u|\tau),\\
&&U(u)= \sum_{i=0}^3 \ell_i(\ell_i+b)\wp (u-\omega_i)
,  ~~~~~~ \ell_i=(\frac{1}{b}-2\alpha_i).
\end{eqnarray*}
When $\ell_i=b l_i$, 
via the 
transformation technique advocated within 
\eqref{eq1}--\eqref{eq7}, one 
realizes that up to a prefactor 
$\lim_{b\to 0}{\cal B}_5 \sim \Phi^{cl}(u|\tau)$ does 
obey Heun equation.

Remarkably, in \cite{Eguchi:1986sb, Fateev:2009aw} it was shown that 
toric two-point conformal blocks under $b\to 0$ 
satisfies Lam\'{e} ($s_1=s_2=s_3=0$) equation to which Heun can 
reduce, i.e.  
\begin{eqnarray}
\label{he}
&&\big\langle V_{-\frac{b}2}(z) V_\alpha(0)\big\rangle_\tau
=\Theta_1(z)^\frac{b^2}{2}
 \eta(\tau)^{2\Delta_\alpha-1-2b^2}
 \Psi(z|\tau),\nonumber\\
&&\Big(
-{b^{-2}}\partial_z^2+{\alpha^2}\wp(z)\Big)
\Psi^{cl}(z|\tau)
=\frac{2i}{\pi}\partial_\tau \Psi^{cl}(z|\tau), ~~~~~~~~
{b}\to0
, ~~~~~~~~b\alpha:~\text{finite}.
\end{eqnarray}
Upon performing 
the WKB  method onto these Schr\"{o}dinger-like equations, 
one is capable of conjecturing the equivalence between 
spheric four-point and toric one-point BPZ conformal 
blocks (without $\Phi_{2,1}$) under special momentum 
assignments just as argued around (3.29) 
in \cite{Fateev:2009me} (see also \cite{Poghossian:2009mk, Hadasz:2009sw}). 
Certainly, along the line of AGT conjecture, 
relating these pure Liouville stuffs to 
elliptic/non-elliptic $\N=2$ $SU(2)\times SU(2)$ 
SCFTs and studying their Nekrasov 
instanton partition functions 
with surface operators inserted under NS limit analogous to 
\cite{Alday:2010vg} remain good future 
issues.

\subsection{Sutherland model}

Let us pause momentarily to consider how 
eCM can be brought to Sutherland 
(or trigonometric Calogero-Sutherland) model. 
In fact, this is quite straightforward, i.e. 
given eCM Hamiltonian 
\begin{eqnarray}
H_E\equiv \frac{1}{4\pi^2}\left(
-\sum_{i=1}^N\frac{\partial^2}{\partial x_i^2}
+2l(l+1) \sum_{1\le i<j \le N}\wp(x_i-x_j)\right)
\end{eqnarray}
where $l$ is the coupling constant 
and the fundamental 
periodicity of 
Weierstrass $\wp$-function is 
$(1,\tau)$, via 
$\tau \rightarrow i\infty$ or 
$q=\exp(\pi i\tau)\to 0$ one obtains from 
$H_E$ 
\begin{eqnarray}
\label{sm}
H_S\equiv \frac{1}{2}\left(
-\sum_{i=1}^N\frac{\partial^2}{\partial x_i^2}+
\sum_{1\le i<j \le N}
\frac{2\beta(\beta-1)\pi^2}{\sin^2 \pi(x_i-x_j)}\right).
\end{eqnarray}

In order to compute wave functions of 
$H_S$ we redefine it by means of the ground state 
$\Delta (X)$ $(X_i = e^{2\pi i x_i})$ as 
\begin{eqnarray*}
H_0\equiv \Delta(X)^{-1}(H_S-e_0)\Delta(X), ~~~~~~~~~~
\Delta(X)\equiv \Big( \frac{1}{\pi} \prod_{i<j} \sin \pi (x_i-x_j)\Big)^{\beta}
\end{eqnarray*}
where $e_0$ comes from $H_S\Delta(X)=e_0\Delta(X)$ and 
$\Delta(X)$ reduces to simply an unitary Vandermonde 
at $\beta=1$. Now, a well-known fact is that 
$H_0$ owns symmetric Jack polynomials 
as its excited eigenstates ($\lambda$: Young tableaux)
\begin{eqnarray}
\label{h0}
H_0J_\lambda(X)=E_\lambda J_\lambda(X)
\end{eqnarray}
which reduce to Schur polynomials at $\beta=1$. 
As pointed out in \cite{Awata:1995ky}, 
a collective chiral boson description 
responsible for $J_\lambda (X)$ exists and its 
Virasoro central charge is related to $\beta$ by 
$c=1-\dfrac{6(1-\beta)^2}{\beta}$. We will return to 
this aspect soon.

\subsection{Gegenbauer versus ${\cal T}_{0,3}(A_1)$}
While a gauge-transformed $H_S$ in \eqref{sm} gets 
reduced to a two-body $A_1$-system, it is widely known that 
its eigenstates become Gegenbauer polynomials: 
\begin{eqnarray}
\label{gg}
C_n^{\nu}(x)=\frac{\Gamma(n+2\nu)}{\Gamma(2\nu)n!}\cdot 
{}_2F_1(-n,n+2\nu,\nu+\frac{1}{2};\frac{1-X}{2}). 
\end{eqnarray}
Though $C_n^{\nu}(x)$ has to be 
viewed as some $\lim_{b\to0} {\cal B}$ as stressed, 
it nevertheless can stand for a 
$b=i$ (4D physical limit) solution to 
\eqref{JJ} being exact w.r.t. $b$ 
whenever $\beta$ in \eqref{sm} 
gets necessarily renormalized to, say, 
$\tilde{\beta}$. This special value of $b$ 
will amount to facilitating our comparison with 
``unrefined" information encoded in ordinary toric diagrams.

In the presence of ${}_2F_1(\cdots)$ in 
\eqref{gg}, we legally doubt 
whether it has something to do with the spheric 
${\cal B}_4(b=i)$ having one $\Phi_{2,1}$ insertion. 
Usually, $\big\langle V_{-\frac{b}2}(x) V_1(0)V_2(1)V_3(\infty)
\big\rangle \equiv {}_2F_1(A,B,C;x)$ 
where 
\begin{eqnarray}
\label{abcn}
\begin{cases}
A=-N,\\
B=\dfrac{1}{\beta}(-\dfrac{2\alpha_1}{\epsilon_1}
-\dfrac{2\alpha_2}{\epsilon_1}+2) +N-1,\\
C=\dfrac{1}{\beta}(-\dfrac{2\alpha_1}{\epsilon_1}+1),\\
N=-\epsilon_1(\alpha_1+\alpha_2-\alpha_3-\frac{b}{2}),~~~~~
\beta=-\dfrac{\epsilon_2}{\epsilon_1}, ~~~~~\epsilon_1=b, ~~~~~
\epsilon_2=\dfrac{1}{b}.
\end{cases}
\end{eqnarray} 
Henceforth, that 
arguments inside ${}_2F_1(\cdots;x)$ of 
$C_n^{\nu}(x)$ are not all independent tells us 
that only two of three 
${\alpha_i}$'s are independent. Still, note that 
$X=\exp(2\pi i x)\in{\mathbb C}^\ast$ makes 
the periodicity of $x\in{\mathbb C}$ even 
explicit. Let us render a justification in the 
following paragraph. 
\begin{figure}[htbp]
 \begin{center}
  \includegraphics[width=120mm]{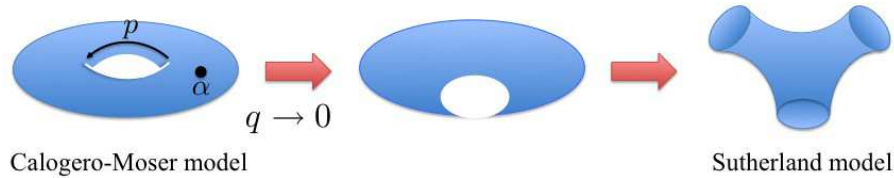}
 \end{center}
 \caption{Carrying out the limit $q\to0$ brings one 
from Calogero-Moser to Sutherland. Note that weights of 
three punctures are not independent because only two 
parameters $(\alpha ,p)$ (converted into ${\cal N}=2^\ast$ $SU(2)$ Coulomb branch 
parameter $\bf{a}$ and adjoint hypermultiplet mass $\bf{m}$ via AGT dictionary) are present.}
 \label{fig:one}
\end{figure}

As shown around \eqref{he}, we manage to think of \eqref{gg} as a 
toric ${\cal B}_2(b=i)$ having 
one $\Phi_{2,1}$ insertion defined on 
a pinched torus as explained in Figure \ref{fig:one}. On the other hand, 
by adhering to \cite{Alday:2009aq, Gaiotto:2009we}, this kind of Riemann surface can also correspond to the $physical$ 
4D ${\cal T}_{0,3}(A_1)$ theory of four free hypermultiplets 
whose bare masses are evaluated by 
$\alpha_1 \pm \alpha_2 \pm \alpha_3$. However, 
because this ${\cal T}'_{0,3}(A_1)$ is yielded via 
a degenerate limit of 
${\cal T}_{1,1}(A_1)$ (that is why a prime is added onto ${\cal T}$) there must exist certain constraint 
between $\alpha_i$'s. 
Combined with \eqref{gg} and \eqref{abcn}, 
one easily arrives at $\alpha_1=\alpha_2$. 
Let us see whether this is consistent with 
what is read off from a toric diagram associated with 
a Calabi-Yau three-fold 
engineering the physical $\N=2^{\ast}$ $SU(2)$ theory at 
$\epsilon_1+\epsilon_2=0$ (or $b=i$). 
\begin{figure}[htbp]
 \begin{center}
  \includegraphics[width=120mm]{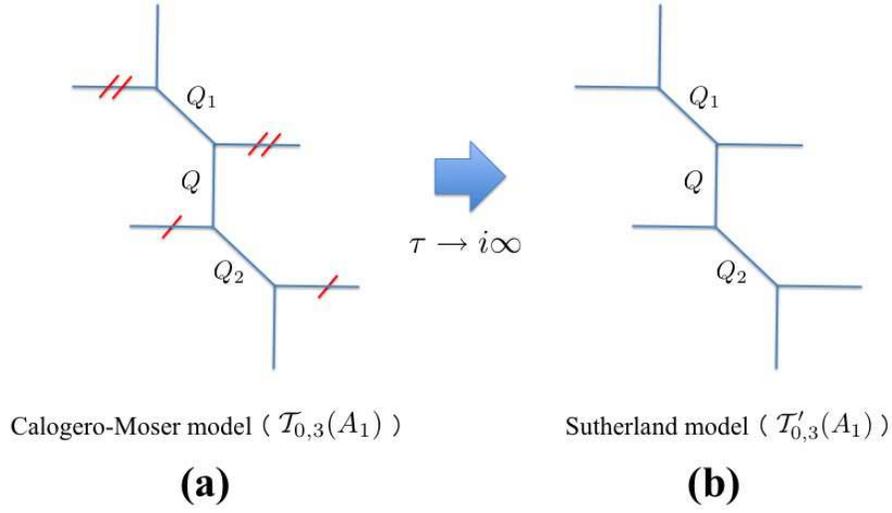}
 \end{center}
 \caption{Transition from ${\cal T}_{0,3}(A_1)$ to 
${\cal T}'_{0,3}(A_1)$ through $\tau\to i\infty$. }
 \label{fig2}
\end{figure}
Based on Figure \ref{fig2}, 
one is able to measure masses 
of four hypermultiplets $m_i$ $(i=1,2,3,4)$ in terms of the distance (or area) 
of blown-up parts of the RHS web diagram $\bf{(b)}$. 
Note also that the correspondence between $\bf{(b)}$ and 
${\cal T}'_{0,3}(A_1)$ (under the name 
$\tilde{T_2}$ strip) 
has been beautifully established in 
\cite{Kozcaz:2010af}. 
In general, 
these areas are basically five-dimensional, i.e. 
$Q_i=\exp (-R {\bf m}_i)$ and $Q=\exp (-R \bf{a})$ which 
fall into 
four-dimensional quantities when $R\to0$ ($R$: size of M-circle). 
Accompanied by some necessarily constant shifts, 
one finds finally that $(m_1, m_2,m_3,m_4)\equiv \alpha_1 \pm \alpha_2 \pm \alpha_3$ subject to $\alpha_1 =\alpha_2$ is fulfilled in Figure \ref{fig2} 
due to $Q_1=Q_2$ (${\bf m}_1={\bf m}_2$) in $\N=2^{\ast}$ cases.

\subsection{Bubbling pants}
Symmetric Jack polynomials with multiple 
variables show up, given the following $k$+3 point 
Liouville conformal block \cite{Kozcaz:2010af, Kane}:  
\begin{eqnarray}
\label{kc}
\frac{\langle \alpha_1|V_2(1) \prod_{i=1}^k V_{-\frac{b}{2}}(z_i) 
|\alpha_3+k\frac{b}{2}\rangle}{\langle \alpha_1|V_2(1) 
|\alpha_3+k\frac{b}{2}\rangle}
\end{eqnarray}
which up to a prefactor 
$\prod_{i=1}^k 
z_i^{b\alpha_1} (1-z_i)^{b\alpha_2}$ is expressed in terms of a 
generalized hypergeometric function ($\gamma>0$): 
\begin{eqnarray*}
_pF^{(\gamma)}_q(a_1,\cdots, a_p; b_1,\cdots, b_q;
z_1,\cdots,z_k)=\sum_{\kappa} 
\frac{1}{|\xi|!}\cdot 
\frac{(a_1)_\kappa^{(\gamma)}\cdots (a_p)_\kappa^{(\gamma)}}{(b_1)_\kappa^{(\gamma)}\cdots (b_q)_\kappa^{(\gamma)}}
\cdot C_\kappa^{(\gamma)}(z_1,\cdots,z_k)
\end{eqnarray*}
with $(p,q)=(2,1)$ and 
\begin{eqnarray*}
a_1=b(\alpha_1 + \alpha_2 - \alpha_3), ~~~~~~~~
a_2 = b(\alpha_1 + \alpha_2 + \alpha_3 +Q), ~~~~~~~~
b_1=2b\alpha_1. 
\end{eqnarray*}
Here, 
$\kappa$ denotes a random partition containing at most 
$k$ rows, 
$(a_i)^{(\gamma)}_\kappa$ the generalized Pochhammer 
symbol and $C_\kappa^{(\gamma)}(z_1,\cdots,z_k)$ 
the $C$-normalized 
Jack polynomial obeying 
\begin{eqnarray*}
\sum_{|\kappa|=\xi}  C_\kappa^{(\gamma)}(z_1,\cdots,z_k)
=\sigma_1(z_i)^{\kappa}, 
~~~~~~~~~~\sigma_n (z_i) =\sum_{i_1 <\cdots <i_n}z_{i_1}\cdots z_{i_n}.
\end{eqnarray*}
Whenever we view $ C_\kappa^{(\gamma)}(z_1,\cdots,z_k)$ as 
the eigenstate of $H_0$ in a $k$-body Sutherland model, 
$\gamma$ denotes 
some continuous parameter depending on $\beta$. 
Three momenta $\alpha_i$'s at $b=i$ must also 
be properly constraint in a fashion 
mentioned before. 
Likewise, in 
a hermitian matrix model $Z$ for instance, when the tree-level potential of 
$Z$ is Gaussian ($M$: $N$$\times$$N$ matrix) one has 
\begin{eqnarray*}
{\cal K} (z_1,\cdots,z_k)\equiv \langle \prod_{i=1}^k 
\det(z_i-M) \rangle \simeq 
\dfrac{\det \big(H_{N+j-1}(z_i)\big)}{\Delta(z)}, ~~~~~~~~~~H_N(z):~ \text{orthogonal polynomial}
\end{eqnarray*}
where $\Delta(z)$ is the usual Vandermonde. Needless to say, 
${\cal K}$ is an eigenstate of a $k$-body 
gauge-transformed Hamiltonian $H_g$
\begin{eqnarray*}
H_g =\Delta(z)^{-1} H_h \Delta(z), 
~~~~~~~~~~H_h=\sum_{i=1}^{k} -\frac{\p^2}{\p z_i^2} + z_i^2.
\end{eqnarray*}
Now we are in a position to ask what is the 
physical content of the situation: 
${\beta}=1$ such that 
Jack reduces to Schur. 
\begin{figure}[htbp]
 \begin{center}
  \includegraphics[width=120mm]{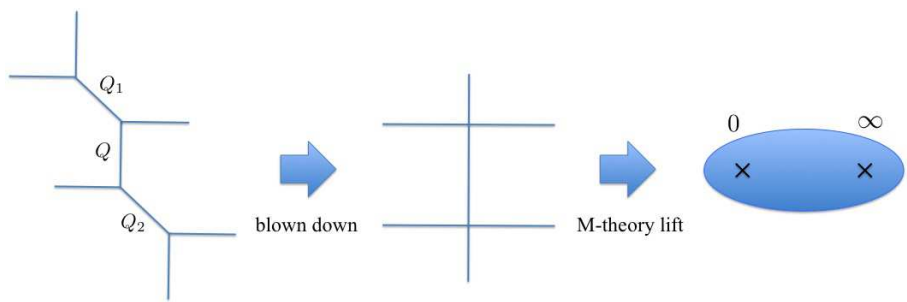}
 \end{center}
 \caption{A blown-down procedure leads to a singular 
toric diagram. }
 \label{fig3}
\end{figure}

\begin{figure}[htbp]
 \begin{center}
  \includegraphics[width=100mm]{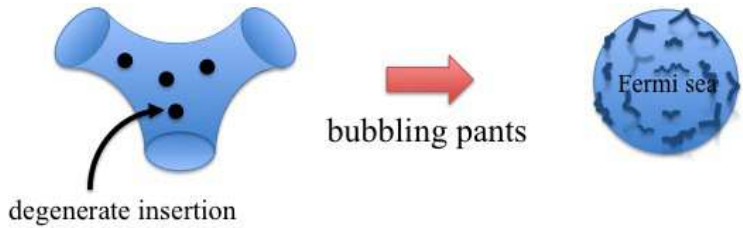}
 \end{center}
 \caption{Schematic picture of bubbling pants.}
 \label{fig4}
\end{figure}
This time we try to resort to Figure \ref{fig3} where $(Q_1, Q_2)$ are blown down 
to be zero-sized. 
This well echoes the fact that the critical 
value ${\beta}=1$ is 
just responsible for ${\bf m}=0$ 
(zero adjoint hypermultiplet mass) in view of 
\eqref{he} and \eqref{sm}. 
Moreover, in M-theory 
the central picture in Figure \ref{fig3} is further lifted to a genus-zero Riemann surface punctured at 
$(0,\infty)$. More explicitly, 
one can think of this singular 
web diagram as what engineers $\N=2$ 
$N_f=2N_c=2$ theory with zero flavor masses. 
Its SW curve to some crude extent gained by thickening the web diagram thereby 
looks like 
\begin{eqnarray*}
\omega +\frac{1}{\omega}=\frac{P_1(v)}{\sqrt{Q_2(v)}}
\end{eqnarray*}
representing a two-punctured Riemann sphere. 
Note that the subscript denotes the degree of 
a monic polynomial. 
Henceforth, that 
many degenerate $V_{-\frac{i}{2}}$ insertions at 
${\beta}=1$ (RHS of Figure \ref{fig3}) are 
captured by Schur polynomials \cite{Awata:1995ky} 
now gets mapped to many 
tachyon excitations inserted on the asymptotic region of the 
self-dual $c=1$ Fermi liquid%
\footnote{For the sake of brevity, we will omit the word 
``self-dual".}. The reason is twofold. 
First, the asymptotic 
collective field describing the 
shape of $c=1$ Fermi surface 
is fermionized to precisely Schur polynomials defined by 
($\lambda$: Young tableaux, $\lambda_i$: $i$-th row length)%
\footnote{See \cite{Tai:2007vc, Tai:2007er} for detailed discussions about 
$c=1$ string theory, Imbimbo-Mukhi type matrix model and Schur polynomials. In particular, 
in \cite{Tai:2007vc} $\N=4$ half-BPS correlators 
independent of Yang-Mills coupling constant $g_{YM}$ 
are shown to coincide with $c=1$ 
tachyon 
scattering amplitudes to all genera. This fact 
may strengthen that 
the blown-down 
${\cal T}'_{0,3}(A_1)$ 
corresponds to $\N=4$ Yang-Mills with $g_{YM}\to 0$. } 
\begin{eqnarray*}
\langle \lambda|Z \rangle=
\frac{\det z^{\lambda_i + k-i}_j}{\det z^{k-i}_j}, 
~~~~~~~~~~i,j=1,\cdots,k.
\end{eqnarray*}
Second, the undisturbed $c=1$ Fermi liquid in the phase space is eventually mirrored to a two-punctured sphere as asserted in \cite{Aganagic:2003qj} via 
the topological B-model language. 
Now we have obtained from ``bubbling pants" picture 
(Figure 
\ref{fig4}) the celebrated free-fermionic (Schur) nature of $c=1$ non-critical 
string theory. Namely, without the 
$massless$ limit of ${\cal T}'_{0,3}(A_1)$ we might never discover 
a description of the old $c=1$ story in terms of 
Sutherland model. 

\section{Summary}
Let us summarize the main idea 
we are after in this article by a chart attached below. 
\begin{figure}[ht]
 \begin{center}
  \includegraphics[width=145mm]{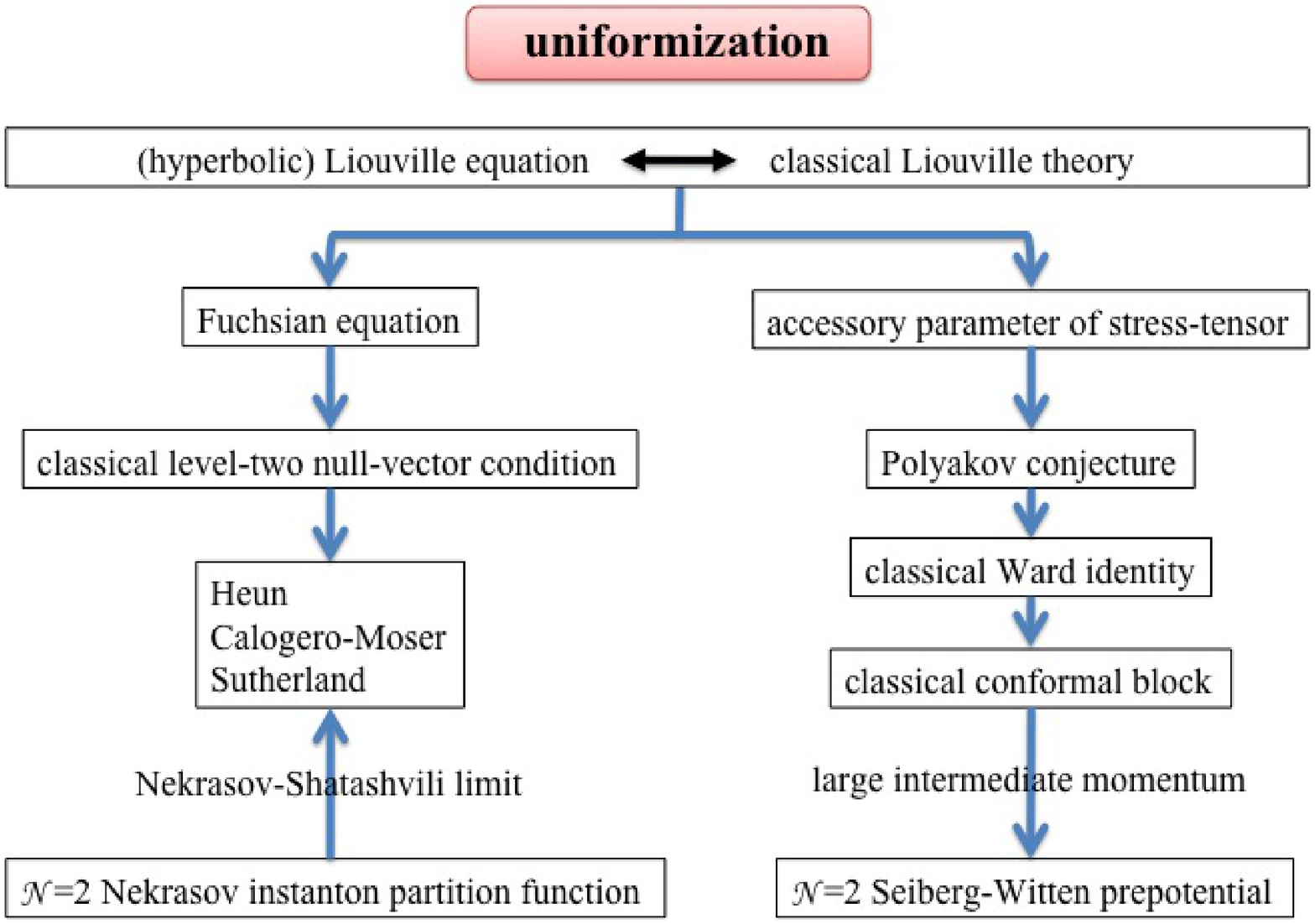}
 \end{center}
\end{figure}

\section*{Acknowledgements} 
TST thanks Takeo Inami, 
Hirotaka Irie, Tetsuji Kimura, Kazuyoshi Maruyoshi, 
Yutaka Matsuo, 
Akitsugu Miwa, Shotaro Shiba, Masato Taki and 
Akihiro Tsuchiya for helpful communications. 
He owes very much Hideaki Iida for his 
discussions and numerous graphical supports. 
He is also grateful to organizers and participants 
of ``3rd Mini Workshop on String Theory" held at KEK. 
TST is supported in part by the postdoctoral program at RIKEN.

\end{document}